\documentclass[twocolumn]{article}

\usepackage{amsmath}
\usepackage{braket}
\usepackage{quantikz}
\usepackage{tikz}
\usepackage{bm}
\usepackage{float}
\usepackage{placeins}
\usepackage[super,sort,compress]{cite}
\usepackage[verbose,hypertexnames=false]{hyperref}

\hypersetup{
colorlinks=false,
allbordercolors=blue,
pdfborderstyle={/S/U/W 1}
}

\begin{document}

\markboth{Khan and Faryad}{State Teleportation using Noisy Quantum Gates}


\title{Analysis of State Teleportation using Noisy Quantum Gates}

\author{Imama Tul Birrah Khan and Muhammad Faryad\footnote{Corresponding author: muhammad.faryad@lums.edu.pk}
\\Department of Physics, Lahore University of Management Sciences, Lahore 54792, Pakistan}

\maketitle

\begin{abstract}
Noise is a major challenge in quantum computing, affecting the reliability of quantum protocols. In this work, we analytically study the impact of various noise processes, such as depolarization, bit flip, and phase flip, on the quantum state teleportation protocol. Each noise process is modeled as a quantum channel and is applied individually to all qubits after the corresponding unitary operations to simulate realistic conditions. We evaluate the fidelity between the ideal and noisy teleported states to quantify the effect of noise. Our analysis shows that the fidelity decreases polynomially, in general, as the noise strength increases for all noise types, highlighting the sensitivity of state teleportation to different noise mechanisms. However, in the low noise regime, the fidelity decreases only linearly, indicating the robustness of the teleportation protocol. These results provide insight into error characterization and can inform strategies for noise mitigation in practical quantum computing applications.
\end{abstract}



\section{Introduction}
\label{sec:level1}

Qubits are fundamental units of quantum computation, but they are highly susceptible to interactions with their environment, leading to decoherence and loss of information. This decoherence decreases the purity of quantum states, posing a significant challenge to performing reliable quantum operations. In quantum teleportation, noise affects the purity of the transmitted state, reducing the efficiency of the protocol \cite{Nielsen:00}. In this work, we study how depolarizing, bit flip, and phase flip noise impact the efficiency of quantum state teleportation, providing insight into the robustness of the protocol in realistic settings.

Quantum communication plays a crucial role in quantum computing, and teleportation protocols have received significant research attention. They are fundamental to quantum information science, enabling the transfer of an unknown quantum state between parties using shared entanglement and classical communication \cite{Nielsen:00,Brassard:96}. Teleportation protocols also help overcome hardware constraints on quantum chips. State teleportation is essential for many quantum technologies, computational capabilities, and advancements \cite{Pirandola:15,Uotila:24}.

Real world quantum systems undergo various noise processes \cite{Resch:21}. Among the noise models used to depict these, depolarizing noise is particularly relevant, as it represents a uniform loss of coherence and information in a quantum system. It is a reasonable approximation because it leads to a completely mixed state with some probability, thereby partially encompassing all Pauli channels frequently used to model noise processes. Its simplicity makes it a practical choice. Moreover, it provides an upper bound on performance, as most other noise models will generally perform worse. The expression for the depolarizing noise channel acting on a bipartite state presented by \cite{Filippov:12} will be used here and generalized to multipartite circuits. In addition to depolarizing noise, we also analytically include bit flip and phase flip noise models, which represent other common types of quantum errors and allow a more comprehensive assessment of teleportation fidelity under realistic conditions.

Other quantum channels have also been used to model the time evolution of qubits under noisy conditions \cite{Georgopoulos:21}. Experimental results on the action of selected local environments on the fidelity of the quantum teleportation protocol, considering realistic and non-ideal entangled resources, have also been reported \cite{Knoll:14}. These include the impact of identical or different types of noise applied partially or completely on the qubits \cite{Fonseca:19}.  {The fragility of entanglement under realistic noise conditions has also been a central challenge in quantum communication. Recent work has demonstrated that advanced encoding strategies, such as hyper entanglement, can significantly enhance noise resilience and enable entanglement distribution even in strongly noisy environments \cite{Im2021}.}

Fidelity between the transmitted and teleported states has been employed as a useful tool to study noise \cite{Kumar:03}. The analysis showed that certain types of decoherence degrade fidelity to classical limits, while others allow quantum advantage to be maintained. The fidelity expressions and evaluation provide a direct comparison with existing results and a framework for assessing different noise conditions \cite{Oh:02,Wilde:13}.  { Related analytical studies have examined the impact of multiple noisy quantum channels on quantum states by evaluating fidelity and coherence degradation \cite{Dutta2023}.}

{Unlike many prior studies that examine isolated noise effects, this work provides a unified analytical framework comparing multiple independent noise channels applied sequentially within a realistic teleportation circuit. This approach enables a direct assessment of how cumulative gate level noise impacts teleportation fidelity.} The model employed here not only considers the gates to be noisy, but also accounts for the noise qubits might encounter in the time between {these} applications of successive gates. The paper is structured to present the noiseless circuits and the analytical development of the depolarizing, bit flip, and phase flip noise models for a multiple-qubit system in Section~\ref{sec:level2}. Following this, in Section~\ref{sec:level3} we analyze the impact of these noise models on a three qubit state teleportation system. Finally, we discuss and conclude our study in Section~\ref{sec:level5}. The findings provide insights into the limitations of current quantum technologies and contribute to the development of error mitigation strategies for quantum communication and computation.

\section{Preliminaries}
\label{sec:level2}

In this section we briefly present the Pauli operators frequently used in noise modeling, pure state teleportation circuit, the noise model, and the expression for fidelity used to assess the impact of the proposed model. In this work, noise is applied after every gate to model realistic {noisy intermediate-scale quantum} (NISQ) era hardware, where errors arise not only from imperfect gate implementations but also from decoherence during idle times between operations. Applying the noise channel after each gate therefore captures both gate induced errors and background environmental noise, providing a more accurate representation of the cumulative noise experienced in practical quantum circuits.

\subsection{Pauli Operators}

{

The Pauli operators play a central role in describing quantum noise channels used throughout this work. Their representation in Dirac notation is given by

\begin{equation}
X = |0\rangle\langle1| + |1\rangle\langle0|,
\end{equation}

\begin{equation}
Y = i|1\rangle\langle0| - i|0\rangle\langle1|,
\end{equation}

\begin{equation}
Z = |0\rangle\langle0| - |1\rangle\langle1|.
\end{equation}
These operators correspond, respectively, to bit flip, bit-phase flip, and phase flip transformations, and are used in constructing the noise models analyzed in the following sections.
}

\subsection{Depolarizing Noise Model in Multi-Qubit Systems}

In real-world quantum systems, noise inevitably affects the teleportation process, leading to imperfect fidelity in the output state. One of the most common types of noise encountered in quantum systems is depolarizing noise, which acts on a single qubit as \cite{Nielsen:00}

\begin{equation}
\mathcal{E}^{\otimes 1}(\rho)
= (1 - p)\rho + \frac{p}{2} I,
\end{equation}

where $\rho$ is the input quantum state, $p$ represents the noise probability, and $I$ is the two-dimensional identity matrix. The depolarizing channel models a loss of quantum information: with probability $1-p$, the state remains unchanged, while with probability $p$, it transforms into the maximally mixed state $I/2$.

Extending the depolarizing channel to a multi-qubit system, where noise is applied independently to each qubit and may occur after all quantum gates involved, the joint noise channel for two qubits was expressed in Ref.~\cite{Filippov:12}. Following the same approach, the effective combined depolarizing model for a three-qubit circuit can be written as

\begin{equation}
\begin{gathered}
\mathcal{E}^{\otimes 3}
= \mathcal{E}^{\otimes 1} \otimes
  \mathcal{E}^{\otimes 1} \otimes
  \mathcal{E}^{\otimes 1}, \\[4pt]
\mathcal{E}^{\otimes 3}(\rho)
= (1-p)^3 \rho
+ \frac{p(1-p)^2}{2}
\bigl(
\rho_1' \otimes \rho_2' \otimes I
+ \rho_1' \otimes I \otimes \rho_3' \\
+ I \otimes \rho_2' \otimes \rho_3'
\bigr)
+ \frac{p^2(1-p)}{4}
\bigl(
\rho_1' \otimes I \otimes I
+ I \otimes \rho_2' \otimes I \\
+ I \otimes I \otimes \rho_3'
\bigr)
+ \frac{p^3}{8}
(I \otimes I \otimes I),
\end{gathered}
\end{equation}

where $\rho_i'$ denotes the single-qubit reduced density matrices of the three-qubit state $\rho$, defined as
$\rho_1' = \mathrm{Tr}_{23}(\rho)$,
$\rho_2' = \mathrm{Tr}_{13}(\rho)$,
and
$\rho_3' = \mathrm{Tr}_{12}(\rho)$.
Incorporating this depolarizing noise modifies each term of the state due to independent decoherence acting on every qubit, resulting in a physically realistic noisy output.

For the general $n$-qubit case, the application of depolarizing noise to each qubit takes the form

\begin{equation}
\mathcal{E}^{\otimes n}(\rho)
=
\bigotimes_{i=1}^{n}
\left(
(1-p)\rho_i'
+
p\,\frac{I}{2}
\right).
\end{equation}

This formulation provides a framework for analyzing how depolarizing noise degrades the fidelity of the teleported quantum state and enables the study of noise resilience in practical quantum teleportation implementations.

\subsection{Bit Flip Noise Model in Multi-Qubit Systems}

Bit flip noise is another common error channel in quantum systems, where a qubit flips from $|0\rangle$ to $|1\rangle$ or vice versa with probability $p$. For a single qubit, the action of the bit flip channel is given by \cite{Nielsen:00}

\begin{equation}
\mathcal{E}(\rho) = (1 - p)\rho + p\, X \rho X,
\end{equation}

where $X$ is the Pauli-$X$ (bit flip) operator.

For a three-qubit system, assuming independent bit flip noise on each qubit, the combined channel can be expressed as

\begin{equation}
\mathcal{E}^{\otimes 3}(\rho)
= \sum_{i,j,k \in \{0,1\}} p_i p_j p_k \,
(X^i \otimes X^j \otimes X^k)\, \rho \, (X^i \otimes X^j \otimes X^k),
\end{equation}

where $p_0 = 1-p$, $p_1 = p$, and $X^0 = I$, $X^1 = X$. Each term corresponds to a combination of bit flips applied independently to the three qubits, capturing all possible single and multiple-qubit flips.

\subsection{Phase Flip Noise Model in Multi-Qubit Systems}

Phase flip noise represents another common error channel, where the relative phase of a qubit is inverted ($|1\rangle \to -|1\rangle$) with probability $p$. For a single qubit, the phase flip channel acts as \cite{Nielsen:00}

\begin{equation}
\mathcal{E}(\rho) = (1 - p)\rho + p\, Z \rho Z,
\end{equation}

where $Z$ is the Pauli-$Z$ (phase flip) operator.

For a three-qubit system with independent phase flip noise on each qubit, the combined channel is

\begin{equation}
\mathcal{E}^{\otimes 3}(\rho)
= \sum_{i,j,k \in \{0,1\}} p_i p_j p_k \,
(Z^i \otimes Z^j \otimes Z^k)\, \rho \, (Z^i \otimes Z^j \otimes Z^k),
\end{equation}

where $p_0 = 1-p$, $p_1 = p$, and $Z^0 = I$, $Z^1 = Z$. This expression captures all possible independent phase flips on the three qubits, allowing a complete analysis of fidelity under phase flip noise.

\subsection{Fidelity}

To quantify the effect of noise on the final quantum state, we employ the measure of fidelity, which is used to assess the closeness of a noisy quantum state to an ideal state \cite{Knill:00,Oh:02}. Fidelity is {a commonly used metric} for quantifying the difference between the states. It is defined as

\begin{equation}
F = \langle \psi_{\rm in} | \rho_{\rm out} | \psi_{\rm in} \rangle,
\label{eq6}
\end{equation}

where $\rho_{\rm out}$ is the state teleported through all noisy gates and $|\psi_{\rm in}\rangle$ is the state originally transmitted.

\subsection{Pure State Teleportation}

Quantum state teleportation is a key protocol in quantum information science that allows for the transfer of an unknown quantum state between spatially separated parties using pre-shared entanglement and classical communication. To accomplish this, the sender and the receiver arrange a correlated pair of particles. The sender then makes a joint measurement, the classical result of which is sent to the receiver, who can perform corresponding operations to create a replica of the state to be teleported \cite{Bennett:93}. 

\begin{figure}[H]
    \centering
    \[
    \resizebox{0.8\columnwidth}{!}{
    \begin{quantikz}
    \lstick{$\ket{\psi}$} & & & \ctrl{1} & \gate{H} & \meter{} & \setwiretype{c} \wire[d][2]{c} \\
    \lstick{$\ket{0}$} & \gate{H} & \ctrl{1} & \targ{} & & \meter{} \wire[d][1]{c} \\
    \lstick{$\ket{0}$} & & \targ{} & & & \gate{X} & \gate{Z} & \rstick{$\ket{\psi}$}
    \end{quantikz}
    }
    \]
    \caption{Circuit diagram for the ideal three-qubit state teleportation protocol. The input state $|\psi\rangle$ is entangled with an ancillary Bell pair, followed by a sequence of Hadamard and CNOT gates and a final two-qubit measurement. Classical outcomes are used to apply the appropriate correction on the receiver’s qubit, yielding a perfect reconstruction of the input state in the absence of noise \cite{Nielsen:00}.}
    \label{fig:teleport_clean}
\end{figure}
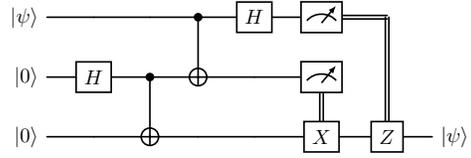

The standard protocol involves an entangled Bell pair shared between two parties. One party holds one part of the entangled pair and the unknown quantum state to be teleported, performs a Bell measurement, and communicates the result to the other, who then applies a corresponding unitary correction to retrieve the original state \cite{Nielsen:00}. The protocol without noise is shown in Fig. \ref{fig:teleport_clean}.

\section{\label{sec:level3}Impact of Noise on State Teleportation}

In this section, we will go through the protocols analytically to work out the teleported output state. Density matrix formalism and Dirac notation will be employed in the analysis.

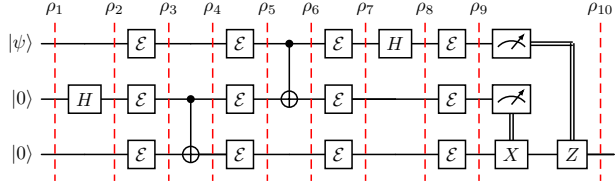
\begin{figure}[t]
    \centering
    \[
    \resizebox{0.51\textwidth}{!}{
    \begin{quantikz}
    \lstick{$\ket{\psi}$}  \slice{$\rho_1$} &  \slice{$\rho_2$} & \gate[wires=1]{\mathcal{E}}  \slice{$\rho_3$} &  \slice{$\rho_4$} & \gate[wires=1]{\mathcal{E}}  \slice{$\rho_5$} & \ctrl{1} \slice{$\rho_6$} & \gate[wires=1]{\mathcal{E}}  \slice{$\rho_7$} & \gate{H}  \slice{$\rho_8$} & \gate[wires=1]{\mathcal{E}}  \slice{$\rho_9$} & \meter{} & \setwiretype{c} \wire[d][2]{c} \slice{$\rho_{10}$}\\
    \lstick{$\ket{0}$}   & \gate{H} & \gate[wires=1]{\mathcal{E}}   & \ctrl{1}  & \gate[wires=1]{\mathcal{E}}   & \targ{}   & \gate[wires=1]{\mathcal{E}}  & \qw &   \gate[wires=1]{\mathcal{E}}   & \meter{}  \wire[d][1]{c} \\
    \lstick{$\ket{0}$}   &   & \gate[wires=1]{\mathcal{E}}   & \targ{} &  \gate[wires=1]{\mathcal{E}}   \qw &&   \gate[wires=1]{\mathcal{E}}  &   & \gate[wires=1]{\mathcal{E}}   & \gate{X} & \gate{Z} & \qw
    \end{quantikz}
    }
    \]
    \caption{Noisy three qubit teleportation circuit used in the analysis. After each unitary operation, a single qubit noise channel $\mathcal{E}$ is independently applied to each qubit to model realistic gate and idle time errors. The density matrices $\rho_1$ through $\rho_{10}$ represent intermediate states at each stage of the protocol. This structure enables analytic evaluation of the cumulative effect of multi-gate noise on the teleported state.}
    \label{fig:3st}
\end{figure}

The state teleportation protocol for a three qubit system with the previously discussed noise model incorporated can be presented as {Fig. \ref{fig:3st}}. 

\vspace{\baselineskip}
Here $\mathcal{E}$ represents the depolarizing channel of a single qubit and $\ket{\psi}$ is the state to be teleported. It can be explicitly written as
\begin{equation}
    \ket{\psi} = \alpha \ket{0} + \beta \ket{1},
\end{equation}
where $\alpha$ and $\beta$ are complex and $|\alpha|^2+|\beta|^2=1$.
Thus, the density matrix for the three-qubit initial state to be teleported can be written as
\begin{equation}    
\begin{gathered}
    \rho_1 = |\alpha|^2 \ket{000}\bra{000} + \alpha\beta^* \ket{000}\bra{100} \\
    + \beta\alpha^* \ket{100}\bra{000} + |\beta|^2 \ket{100}\bra{100},
\end{gathered}
\end{equation}
where $^*$ represents complex conjugates. The Hadamard gate can be applied to a qubit using the expression
\begin{equation}
    H = \frac{1}{\sqrt{2}} \left( \ket{0} \bra{0} + \ket{0} \bra{1} + \ket{1} \bra{0} - \ket{1} \bra{1} \right).
\end{equation}
After the application of the Hadamard gate to the second qubit to create superposition, we get
\begin{equation}
    \rho_2 = (I \otimes H \otimes I) \, \rho_1 \, (I \otimes H \otimes I).
\end{equation}

The application of the depolarizing channel on $\rho_2$ gives
\begin{equation}
    \rho_3 = \mathcal{E}^{\otimes3}(\rho_2).
\end{equation}

Following this, a CNOT gate can be applied between two qubits using the expression
\begin{equation}
    CNOT = \ket{00}\bra{00} + \ket{01}\bra{01} + \ket{10}\bra{11} + \ket{11}\bra{10}.
\end{equation}

The application of CNOT between the second and third qubits returns 
\begin{equation}
    \rho_4 = (I \otimes CNOT) \, \rho_3 \, (I \otimes CNOT).
\end{equation}

Another round of noisy gates evolves $\rho_4$ to
\begin{equation}
    \rho_5 = \mathcal{E}^{\otimes3}(\rho_4).
\end{equation}

This is followed by a CNOT gate between the first two qubits to give
\begin{equation}
    \rho_6 = (CNOT \otimes I) \, \rho_5 \, (CNOT \otimes I).
\end{equation}

The third tier of depolarizing noisy gates leads to
\begin{equation}
    \rho_7 = \mathcal{E}^{\otimes3}(\rho_6).
\end{equation}

Application of the Hadamard gate to the first qubit to produce superposition gives
\begin{equation}
    \rho_8 = (H \otimes I \otimes I) \, \rho_7 \, (H \otimes I \otimes I).
\end{equation}

The last density matrix before measurement, after passing through the final set of noisy gates or being susceptible to random noise in the wait time, comes out as
\begin{equation}
    \rho_9 = \mathcal{E}^{\otimes3}(\rho_8).
\end{equation}

The first two qubits are then measured, and the measurement results lead to a corresponding application of relevant operations on the third qubit, which gives the final states presented below for various noise channels employed.

\subsection{Depolarizing Noise}

The final state after the application of the noise model presented above with $\mathcal{E}$ as depolarizing noise channels is as follows
\begin{equation}
\begin{gathered}
    \rho_{10} = ((1-p)^9 |\alpha|^2 + \frac{1-(1-p)^9}{2}) \ket{0}\bra{0} + \\
    (1-p)^{12} \alpha \beta^* \ket{0}\bra{1} + (1-p)^{12} \alpha^* \beta \ket{1}\bra{0} + \\
    ((1-p)^9 |\beta|^2 + \frac{1-(1-p)^9}{2}) \ket{1}\bra{1}.
\end{gathered}
\end{equation}

Ideally, the  state $\ket{\psi}$ should have been teleported through the state teleportation protocol but due to the noisy gates and vulnerable nature of qubits to random noise, the state  has been altered. It is  dependent on the probability $p$ associated with the depolarizing channels. Setting $p$ equal to zero to draw a parallel with the ideal case, $\ket{\psi}$ is retrieved as it should be. The fidelity between the final density matrix and the input state in that case would also retreat to 1.

Whereas, the fidelity $F$ for the three qubit noisy state teleportation circuit using the expression given in \eqref{eq6} was found to be
\begin{equation}
    F = \bra{\psi} \rho_{10} \ket{\psi}.
\end{equation}

\begin{figure}[H]
    \centering
    \includegraphics[width=\linewidth]{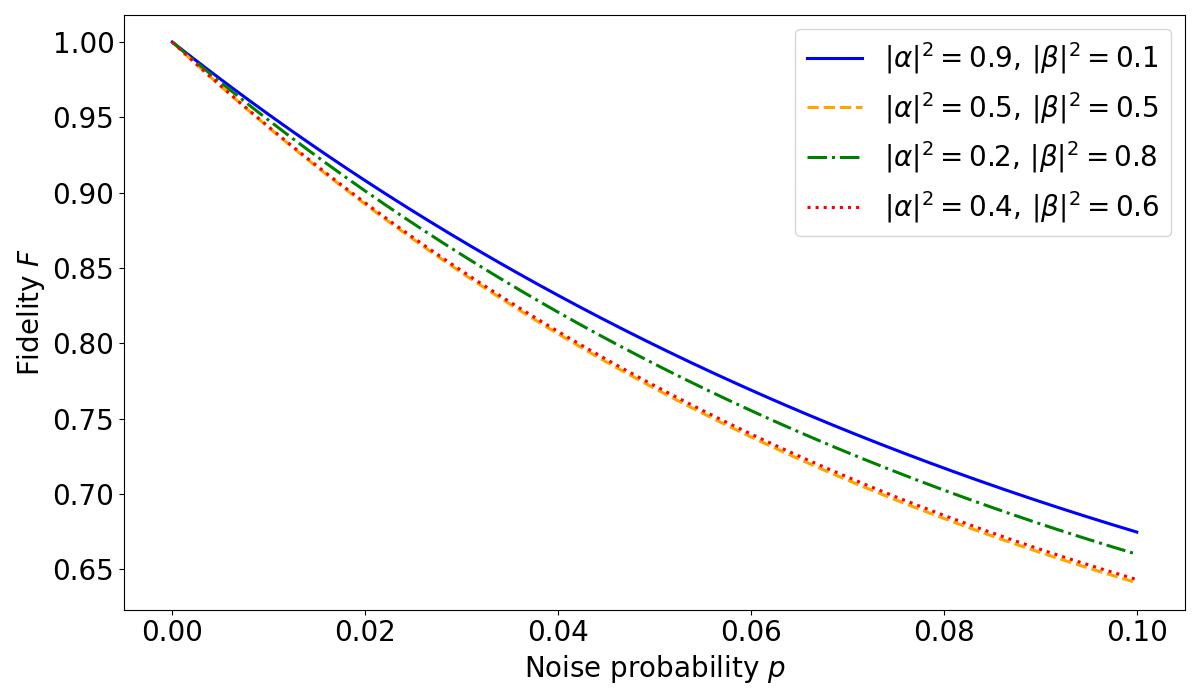}
    \caption{{Fidelity of the teleported state as a function of depolarizing noise probability $p$ for different choices of the input state parameters $(\alpha,\beta)$. The fidelity decreases as the noise strength increases, reflecting the progressive loss of quantum coherence and the convergence of the output state toward a maximally mixed state.}}
    \label{fig:three_qubit_fid}
\end{figure}

The variation of this fidelity with the noise probability $p$ is presented in {Fig. \ref{fig:three_qubit_fid}}. There is a slightly steeper decrease in fidelity when $\alpha$ and $\beta$ are equal (i.e., for an equal superposition of $\ket{0}$ and $\ket{1}$), but it eventually falls to the same minimum. In the worst-case scenario, the output state becomes a maximally mixed single qubit state $I/2$, which has a fidelity of $F = 1/2$. The fidelity decreases rapidly with increasing noise strength and approaches the classical limit of $F = 0.5$ as $p \rightarrow 1$, reflecting the convergence of the output state towards the maximally mixed state. Superposition states ($|\alpha| = |\beta|$) exhibit the steepest decline due to their higher coherence content, which is most susceptible to depolarization.

For small $p$, the fidelity can be approximated as
\begin{equation}
    F \approx 1 - p \left( 6 |\alpha|^2 |\beta|^2 + \frac{9}{2} \right).
\end{equation}

\subsection{Bit-Flip Noise}

The final state after the application and analysis of the noise model presented above with $\mathcal{E}$ as bit flip noise channels is as follows:
\begin{equation}
\begin{aligned}
\rho_{10} ={}& 4 \Big[
(u_1 |\alpha|^2 + u_2 |\beta|^2 + u_3) \ket{0}\bra{0} \\
&+ (u_2 |\alpha|^2 + u_1 |\beta|^2 + u_3) \ket{1}\bra{1} \\
&+ (u_4 \alpha \beta^* + u_5 \beta \alpha^*) \ket{0}\bra{1} \\
&+ (u_5 \alpha \beta^* + u_4 \beta \alpha^*) \ket{1}\bra{0}
\Big],
\end{aligned}
\end{equation}

where the constants $u_i$ depend on the bit flip probability $p$ as:

\begin{align*}
u_1 &= -64p^9 + 288p^8 - 576p^7 + 672p^6 - 504p^5 + 252p^4 \\
&\quad - 84p^3 + \frac{73}{4} p^2 - \frac{5}{2} p + \frac{1}{4},\\
u_2 &= 64p^9 - 288p^8 + 576p^7 - 672p^6 + 504p^5 - 252p^4 \\
&\quad + 84p^3 - \frac{71}{4} p^2 + 2p,\\
u_3 &= -\frac{1}{4} p^2 + \frac{1}{4} p,\\
u_4 &= 32p^{11} - 64p^{10} - 128p^9 + 640p^8 - 1108p^7 + 1120p^6 \\
&\quad - 742p^5 + \frac{1337}{4} p^4 - \frac{205}{2} p^3 + \frac{83}{4} p^2 - \frac{11}{4} p + \frac{1}{4},\\
u_5 &= -32p^{11} + 64p^{10} + 128p^9 - 640p^8 + 1108p^7 - 1120p^6 \\
&\quad + 742p^5 - \frac{1335}{4} p^4 + \frac{405}{4} p^3 - \frac{79}{4} p^2 + 2p.
\end{align*}

The fidelity for the three-qubit teleportation circuit under bit-flip noise can be written as
\begin{equation}
    F = \bra{\psi} \rho_{10} \ket{\psi}.
\end{equation}

\begin{figure}[H]
    \centering
    \includegraphics[width=\linewidth]{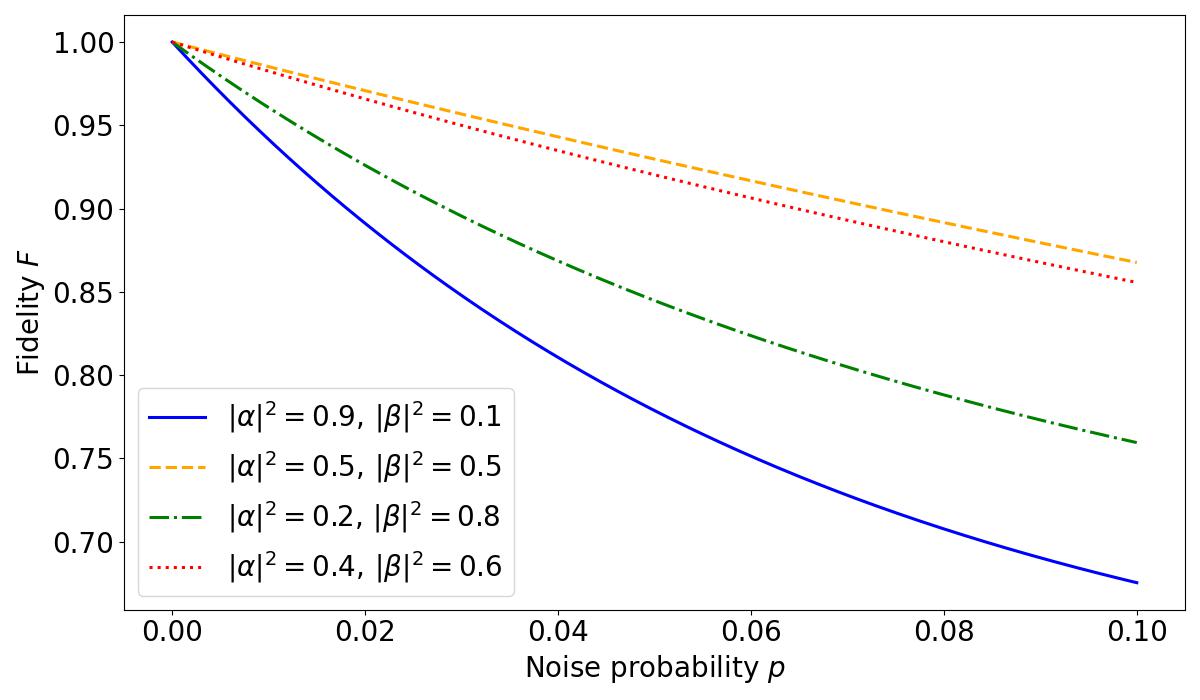}
    \caption{{Fidelity of the teleported state as a function of bit-flip noise probability $p$ for different choices of the input state parameters $(\alpha,\beta)$. The fidelity decreases with increasing noise strength due to population exchange errors introduced by bit flips, reflecting how these errors degrade teleportation performance.}}
    \label{STBit}
\end{figure}

The variation in this fidelity with noise probability is presented in {Fig. \ref{STBit}}. For small values of the bit flip probability $p$, this fidelity can be approximated linearly as
\begin{equation}
    F \approx 1 - p \Big\{9 - 14 |\alpha|^2 |\beta|^2 - 8 \big[(\alpha\beta^*)^2 + (\beta\alpha^*)^2\big]\Big\}.
\end{equation}

The fidelity of the teleported state under bit flip noise decreases as the bit flip probability $p$ increases. When $p=0$, the teleportation is ideal, and the fidelity is restored to $F = 1$. For small $p$, the linear approximation captures the first-order effect of noise, where both population imbalance and coherence terms of the input state contribute to the reduction in fidelity. The fidelity decreases with increasing bit flip probability $p$, but at a slower rate compared to depolarizing and phase flip noise. States biased toward the computational basis ($|\alpha| \approx 1$ or $|\beta| \approx 1$) exhibit stronger degradation, while coherent superpositions are relatively more resilient.

\subsection{Phase-Flip Noise}

The final state after the application and analysis of the noise model presented above with $\mathcal{E}$ as phase flip noise channels is as follows:
\begin{equation}
\rho_{10} = |\alpha|^2  \ket{0}\bra{0} + u_6 \alpha \beta^* \ket{0}\bra{1} + u_6  \beta \alpha^*  \ket{1}\bra{0} + |\beta|^2  \ket{1}\bra{1},
\end{equation}

where the constant $u_6$ depends on the phase flip probability $p$ as
\begin{equation}
\begin{aligned}
u_6 ={}& 256p^8 - 1024p^7 + 1792p^6 - 1792p^5 + 1120p^4 \\
&- 448p^3 + 112p^2 - 16p + 1.
\end{aligned}
\end{equation}

The fidelity of the teleported state is given by
\begin{equation}
F = \bra{\psi}\rho_{10}\ket{\psi}.
\end{equation}

\begin{figure}[H]
\centering
\includegraphics[width=\linewidth]{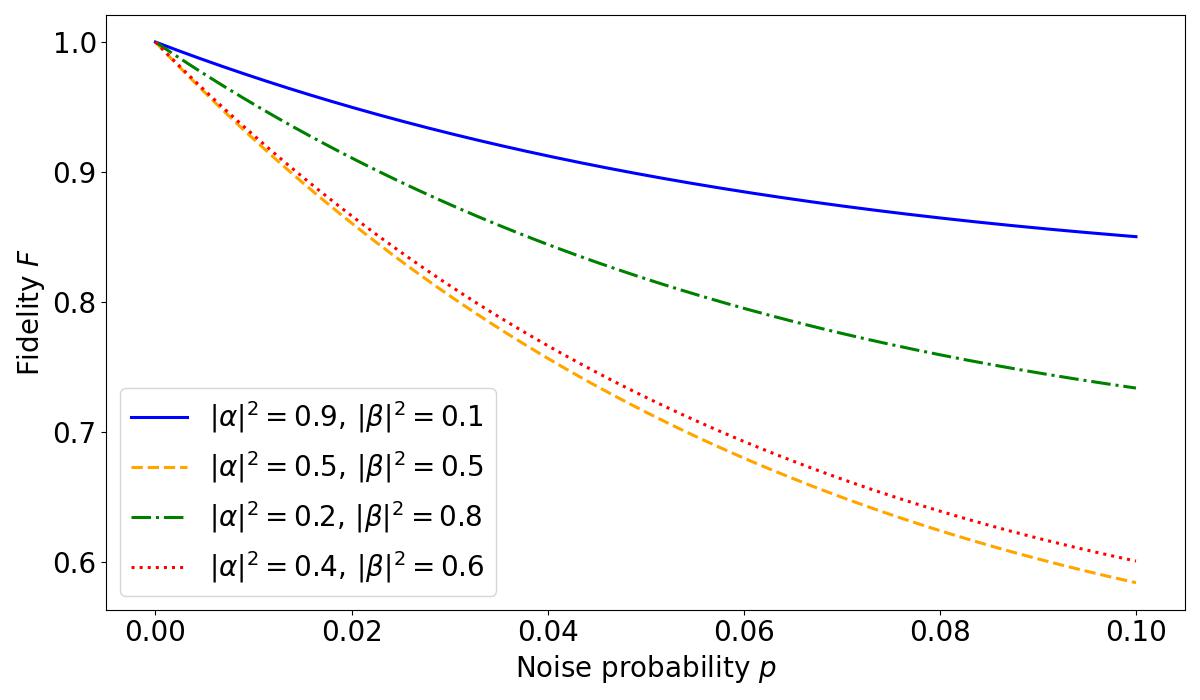}
\caption{{Fidelity of the teleported state as a function of phase-flip noise probability $p$ for different choices of the input state parameters $(\alpha,\beta)$. The fidelity reduction reflects suppression of off-diagonal coherence terms, demonstrating the sensitivity of teleportation to phase errors that directly destroy quantum coherence.}}
\label{STPhase}
\end{figure}

The variation in this fidelity with noise probability is presented in {Fig. \ref{STPhase}}.  The fidelity decreases as the phase flip probability $p$ increases. When $p=0$, the teleportation is ideal, and the fidelity is restored to $F = 1$. States that are equal superpositions of $\ket{0}$ and $\ket{1}$ experience a steeper reduction in fidelity, highlighting the greater sensitivity of coherent superpositions to phase flip errors. Phase flip errors suppress the off-diagonal elements of the density matrix, directly destroying quantum coherence. As a result, states with large coherence ($|\alpha\beta|$) show the fastest fidelity degradation.

For small values of the phase-flip probability $p$, the fidelity can be approximated linearly as
\begin{equation}
F \approx 1 - p \, (32  |\alpha|^2 |\beta|^2).
\end{equation}

\section{\label{sec:level5}Discussion and Conclusions}

The analytical expressions and fidelity plots presented in this work provide a comprehensive understanding of how different forms of quantum noise impact the performance of the three qubit quantum state teleportation protocol. Across all noise models considered, the fidelity exhibits a rapid decline as the noise probability increases. Significant degradation occurs even within the small noise regime ($0 < p < 0.1$), which is characteristic of contemporary NISQ devices. This underscores the strong susceptibility of teleportation circuits to accumulated gate errors and decoherence, especially when noise acts after every gate operation and during idle intervals.

The depolarizing noise model demonstrates the expected uniform degradation of both populations and coherences, driving the teleported state towards the maximally mixed state as $p$ increases. The fidelity curves saturate around $F = 0.5$, consistent with the classical limit for single qubit teleportation. Input states with equal superposition ($|\alpha| = |\beta|$) experience the steepest reduction in fidelity because their coherence content is most strongly affected by depolarizing noise. This behavior highlights that the degree of quantum coherence in the input state directly influences how rapidly fidelity deteriorates under uniform decoherence.

In the bit flip noise model, a distinct pattern emerges. Since bit flips primarily affect the populations of the computational basis states, population imbalanced inputs (e.g., states close to $|0\rangle$ or $|1\rangle$) experience greater fidelity loss. In contrast, coherent superposition states remain comparatively more resilient. This asymmetry reflects the classical nature of bit flip errors and indicates that teleportation is more robust against bit level population noise than against coherence destroying processes.

Phase flip noise proves to be the most detrimental of all noise types considered. It {targets} the off-diagonal components of the density matrix, the protocol’s sensitivity to these errors becomes immediately apparent. States with high coherence ($\alpha\beta \neq 0$), especially equal superpositions, exhibit the most rapid fidelity reduction. The steep decline confirms that the preservation of quantum coherence is the central resource enabling successful teleportation, and phase-flip noise directly compromises this resource more than population altering noise mechanisms, hence compromising fidelity the most.

A consistent theme across all models is that the fidelity loss is state dependent. Superposition states suffer most under coherence destroying noise (depolarizing and phase flip), whereas states aligned with the computational basis are more susceptible to population altering bit flip errors. This state dependent behavior suggests that teleportation may serve not only as a communication protocol but also as a practical diagnostic tool for identifying dominant noise types in quantum hardware by examining fidelity variations across different input states.

{
Although the present analysis focuses on a three qubit teleportation circuit, the qualitative behavior extends to larger multi qubit systems. As the number of qubits and gate operations increases, independent noise channels compound, leading to accelerated fidelity degradation. This cumulative effect arises because each qubit experiences decoherence and operational imperfections repeatedly throughout the protocol. Consequently, teleportation circuits involving more qubits are increasingly sensitive to noise, highlighting the importance of error mitigation and fault-tolerant design strategies for scalable quantum communication architectures.}

Overall, the results clearly demonstrate that realistic noise levels severely limit the performance of multi-gate teleportation circuits. Achieving high fidelity teleportation in practical systems will require substantial improvements in qubit coherence times, gate fidelities, and the integration of error mitigation or encoding strategies. The quantitative characterization provided here offers a benchmark for assessing the robustness of teleportation under realistic noise conditions and serves as a foundation for the development of more noise resilient quantum communication and computation protocols.

\section*{Disclosure}
The authors declare no conflict of interest. Also, no data was generated or used for the research reported in this paper.

\end{document}